\documentclass[12pt]{article}

\begin{document}

\title{Four-Index Equations for Gravitation\\ and the Gravitational Energy-Momentum Tensor \footnote{Published in: ''Z.Zakir (2003) \textit{Structure of Space-Time and Matter}, CTPA, Tashkent.'' }}
\author{Zahid Zakir\thanks{E-mail: zahid@in.edu.uz}\\Centre for Theoretical Physics and Astrophysics,\\ P.O.Box 4412, Tashkent 700000 Uzbekistan}
\date{June 10, 1999;\\
Revised October 17, 2003.}
\maketitle

\begin{abstract}
A new treatment of the gravitational energy on the basis of 4-index
gravitational equations is reviewed. The gravitational energy for the
Schwarzschild field is considered.
\end{abstract}

\section{Introduction}

A covariant physical characteristics of the gravitational field is the
Riemann curvature tensor, and it is natural that the problems with the
energy-momentum of gravitation can be solved if we can express the
gravitational energy in terms of this tensor.

In the papers \cite{Za},\cite{Za1} a new generalized 4-index version of the
Einstein equations with the Riemann tensor has been formulated, and the
local energy-momentum tensors for the system of gravitation field and
matter, linearly depending on the curvature tensor, have been constructed as
4-index tensors.

In the present paper some consequences of this treatment, including the
calculation of the gravitational energy for a mass point, will be presented.

\section{Four-index equations for the gravitational field}

In the standard Einstein-Gilbert gravitational action one can add to the
Ricci tensor or to the Riemann tensor arbitrary functions (tensors) with
zero contractions:

\begin{equation}
R=g^{km}R_{km}=\frac{1}{2}(g^{km}g^{il}-g^{im}g^{kl})(R_{iklm}-\kappa
V_{iklm}),
\end{equation}
where $\kappa =8\pi k$, and $L_{m}$ is the matter Lagrangian, $V_{iklm}$ has
the same symmetry properties as $R_{iklm}$, and $g^{il}V_{iklm}=0$.

So, we can start from the new action function :

\begin{equation}
S=\frac{1}{2}\int d\Omega \sqrt{-g}\left[ \frac{1}{2}%
(g^{km}g^{il}-g^{im}g^{kl})\left( -\frac{1}{\kappa }R_{iklm}+V_{iklm}\right)
+L_{m}\right] ,
\end{equation}
which is fully equivalent to the Einstein-Gilbert action function Eq. (\ref
{EH}){\it .} Then we obtain for the variation of the action function \cite
{Za}:

\begin{equation}
\delta S=-\frac{1}{2}\int d\Omega \sqrt{-g}\delta
g^{km}g^{il}(G_{iklm}-T_{iklm})=0,
\end{equation}
where $T_{iklm}=V_{iklm}+T_{iklm}^{(m)}$, and:
\begin{equation}
G_{iklm}=\frac{1}{\kappa }\left[ R_{iklm}-\frac{1}{(d-1)(d-2)}%
(g_{il}g_{km}-g_{im}g_{kl})R\right] ,
\end{equation}
\[
T_{iklm}^{(m)}=\frac{1}{(d-2)}%
(g_{km}T_{il}-g_{kl}T_{im}+g_{il}T_{km}-g_{im}T_{kl})-
\]

\begin{equation}
-\frac{1}{(d-1)(d-2)}(g_{il}g_{km}-g_{im}g_{kl})T.
\end{equation}
Here $d$ is the spacetime dimensionality, and $T_{iklm}$ has the same
structure as the Riemann tensor having the representation:
\[
R_{iklm}=C_{iklm}+\frac{1}{(d-2)}%
(g_{km}R_{il}-g_{kl}R_{im}+g_{il}R_{km}-g_{im}R_{kl})-
\]

\begin{equation}
-\frac{1}{(d-1)(d-2)}(g_{il}g_{km}-g_{im}g_{kl})R,
\end{equation}
where $C_{iklm}$ is the Weyl tensor with zero contractions $%
g^{il}C_{iklm}=g^{km}C_{iklm}=0$.

Thus, we obtain the equations:

\begin{equation}
g^{il}(G_{iklm}-T_{iklm})=0.
\end{equation}

In a general case the expression in the parenthesis is not equal to zero for
the arbitrary $V_{iklm}$ and we can not simply exclude the contractional
factor $g^{il}$. However the tensor $V_{iklm}$ has 10 independent components
which is equal to the number of the Riemann tensor components in the vacuum (%
$T_{km}^{(m)}=0$) where it is reduced to the Weyl tensor. Therefore, if in
this case we choose the $V_{iklm}$ as equal to:

\begin{equation}
\frac{1}{\kappa }G_{iklm}=\frac{1}{\kappa }C_{iklm}=V_{iklm},  \label{Vac}
\end{equation}
the equations hold identically for the solutions of the Einstein equations.
Thus, we may write the 4-index equations for the gravitational field as \cite
{Za}:

\begin{equation}
G_{iklm}=T_{iklm}.
\end{equation}

We see that $V_{iklm}$ can be considered as the 4-index energy-momentum
density tensor for the gravitational field. Although its 2-index contraction
vanish, in the 4-index form it allows one to determine a nonzero, local and
positive defined energy-momentum tensor for the gravitational field.

The tensors $G_{iklm}$ and $T_{iklm}$ have the symmetry properties of the
Riemann tensor and, therefore, we have 20 equations. The tensor $G_{iklm}$
is a function of the metric tensor $g_{ik}$ which has 6 independent
components. The tensor $T_{iklm}^{(m)}$ has been combined from the ordinary
energy-momentum tensor of the matter $T_{ik}$ and it has 4 independent
functions (the energy density $\epsilon $ and 3 components of the velocity).
These 10 functions obey to 10 Einstein equations. The tensor $V_{iklm}$
gives additional 10 independent components.

So, we have 20 equations for 20 independent functions. If we take solutions
of the Einstein equations for the metrics and $T_{ik}$, then we have the
additional 10 equations for 10 components of $V_{iklm}$. Therefore, the
solutions of the Einstein equations exactly define all components of $%
V_{iklm}$ and in the paper we can find $V_{iklm}$ for the known standard
metrics.

In the vacuum $T_{ik}=T=0,$ $R_{il}=R=0$ and we have the equations Eq.(\ref
{Vac}). We see that in the vacuum the tensor $V_{iklm}$ plays the role of
the source for the empty spacetime curvature $C_{iklm}$.

The covariant derivatives of the 4-index tensors are also related as:

\begin{equation}
G_{klm}^{i}{}_{;i}=T_{klm;i}^{i}.
\end{equation}
In the case $d=4$ we have:
\begin{equation}
G_{.klm;i}^{i}=T_{km;l}-T_{kl;m}-\frac{1}{3}(g_{km}T_{,l}-g_{kl}T_{,m}),
\end{equation}
\begin{equation}
T_{klm;j}^{j(m)}=\frac{1}{2}\left[ T_{km;l}-T_{kl;m}-\frac{1}{3}%
(g_{km}T_{;l}-g_{kl}T_{;m})\right] =\frac{1}{2}G_{.klm;i}^{i}.
\end{equation}
Then we obtain the relationship:

\begin{equation}
V_{klm;j}^{j}=G_{klm;j}^{j}-T_{klm;j}^{j(m)}=\frac{1}{2}G_{.klm;i}^{i}.
\end{equation}
and, therefore, 
\begin{equation}
V_{klm;j}^{j}=T_{klm;j}^{j(m)}
\end{equation}

In the vacuum, therefore, there are local conservation laws:

\begin{equation}
G_{\cdot klm;j}^{j}=V_{klm;j}^{j}=0.
\end{equation}

The integral energy-momentum tensor for the system of matter and
gravitational field can be defined as:

\begin{equation}
P_{lm}^{i}=\int dS_{k}T_{..lm}^{ik}.
\end{equation}

On the hypersurface $x^{0}=const$ we have:

\begin{equation}
P_{.lm}^{k}=\int d^{3}x\sqrt{-g}T_{..lm}^{0k}=\int d^{3}x\sqrt{-g}%
(T_{..lm}^{(m)0k}+V_{..lm}^{0k}).
\end{equation}

The energy-momentum vector for matter can be obtained as: $P^{i}=P_{.k}^{ik}$%
. Finally, the 3-index integral energy-momentum of the gravitational field
can be defined as:

\begin{equation}
P_{.lm}^{(g)i}=\int d^{3}x\sqrt{-g}V_{..lm}^{0i}.
\end{equation}

\section{The gravitational energy for the Schwarzschild field}

Let us consider the energy of the Schwarzschild field with the line element:

\begin{equation}
ds^{2}=\left( 1-\frac{r_{g}}{r}\right) dt^{2}-\frac{dr^{2}}{1-\frac{r_{g}}{r}%
}-r^{2}(d\vartheta ^{2}+\sin ^{2}\vartheta d\varphi ^{2}),
\end{equation}
where $r_{g}=2Gm$ is the gravitational radius, and the components of the
metric are: $g_{22}=-r^{2}$, $g_{33}=-r^{2}\sin ^{2}\vartheta $, and:
\begin{equation}
g_{00}=g_{11}^{-1}=1-\frac{r_{g}}{r}.  \nonumber
\end{equation}
We calculate the energy-momentum tensor:

\begin{equation}
V_{lm}^{ik}=\frac{1}{\kappa }R_{lm}^{ik}
\end{equation}
for this solution of the Einstein equations. Nonzero components of the $%
V_{iklm}=R_{iklm}/\kappa $ with this metric are:
\begin{equation}
V_{0101}=\frac{r_{g}}{\kappa r^{3}}=-V(r)g_{00}g_{11},
\end{equation}
\begin{equation}
V_{0202}=-\frac{r_{g}(r-r_{g})}{2\kappa r^{2}}=\frac{1}{2}V(r)g_{00}g_{22},
\end{equation}
\begin{equation}
V_{0303}=-\frac{r_{g}(r-r_{g})}{2\kappa r^{2}}\sin ^{2}\vartheta =\frac{1}{2}%
V(r)g_{00}g_{33},
\end{equation}
\begin{equation}
V_{1212}=\frac{r_{g}}{2\kappa (r-r_{g})}=\frac{1}{2}V(r)g_{11}g_{22},
\end{equation}
\begin{equation}
V_{1313}=\frac{r_{g}\sin ^{2}\vartheta }{2\kappa (r-r_{g})}=\frac{1}{2}%
V(r)g_{11}g_{33},
\end{equation}

\begin{equation}
V_{2323}=-\frac{r_{g}r}{\kappa }\sin ^{2}\vartheta =-V(r)g_{22}g_{33},
\end{equation}
where:

\begin{equation}
V(r)=\frac{r_{g}}{\kappa r^{3}}=\frac{m}{4\pi r^{3}}=-\frac{m}{8\pi }\frac{%
\partial }{\partial r}(r^{-2}).
\end{equation}

We see, that the 2-index contraction of this tensor vanishes:

\[
V_{il}=g^{km}V_{iklm}=g_{il}[-V(r)+\frac{1}{2}V(r)+\frac{1}{2}V(r)]=0. 
\]

The physical components of the gravitational energy-momentum tensor $%
V_{..lm}^{ik}=g^{ip}g^{kq}V_{pqlm}$ are:
\begin{equation}
V_{..01}^{01}=V_{..10}^{10}=V_{..23}^{23}=V_{..32}^{32}=-V(r),
\end{equation}
\begin{equation}
V_{..02}^{02}=V_{..20}^{20}=V_{..03}^{03}=V_{..30}^{30}=V_{..12}^{12}=
\end{equation}

\begin{equation}
=V_{..21}^{21}=V_{..13}^{13}=V_{..31}^{31}=\frac{1}{2}V(r).
\end{equation}

They allow us to calculate one of components of integral gravitational
energy-momentum of the static mass point as:

\begin{equation}
cP_{.01}^{(g)1}=\int dS_{0}\sqrt{-g}V_{..01}^{01}=\int dS_{0}\sqrt{-g}[-V(r)%
].
\end{equation}
The spatial volume integral can be represented as a spatial surface
integral, and we obtain:
\begin{equation}
cP_{.01}^{(g)1}=\frac{m}{8\pi }\int dS_{0}\sqrt{-g}\frac{\partial }{\partial
r}(r^{-2})=\frac{m}{8\pi }\int df_{0r}r^{-2}=\frac{1}{2}n_{r}m,
\end{equation}
where $df_{0r}=n_{r}r^{2}do$ is 2-dimensional surface element with the
normal vector $n_{r}$ along $r$.

\end{document}